\documentclass[11pt,a4paper]{article}
\usepackage{authblk}
\usepackage{epsfig}
\usepackage{subfigure}
\usepackage{amsmath}
\usepackage{amssymb}
\usepackage{mathtools}
\usepackage{graphicx}
\graphicspath{{plots/}}
\usepackage{hyperref}
\usepackage{csquotes}
\usepackage{multirow}
\usepackage[left=.45in,right=.45in,top=.6in,bottom=.6in,headheight=14.5pt]{geometry}
\usepackage{multirow}
\usepackage{xcolor}
\usepackage{pdflscape}
\usepackage[title]{appendix}
\usepackage[left=.45in,right=.45in,top=.6in,bottom=.6in,headheight=14.5pt]{geometry}
\usepackage{fmtcount} 
\usepackage{array,multirow}
\newcolumntype{C}[1]{>{\centering\arraybackslash}m{#1}}
\usepackage{epsfig}
\usepackage{float}
\usepackage{geometry}
\usepackage{amsmath}
\usepackage{cite}
\usepackage{ctable}

\newgeometry{vmargin={10mm,20mm},hmargin={19mm,19mm}}

\title{Neutrino Phenomenology in a Model with Generalized CP symmetry within Type-I seesaw framework}

\author[1]{Tapender\thanks{tapenderphy@gmail.com}}
\author[2]{Sanjeev Kumar\thanks{ skverma@physics.du.ac.in}}
\author[1] {Surender Verma\thanks{s\_7verma@hpcu.ac.in, Corresponding Author}}
\affil[1]{Department of Physics and Astronomical Science, Central University of Himachal Pradesh, Dharamshala-176215, INDIA}
\affil[2]{Department of Physics and Astrophysics, University of Delhi, Delhi-110007, INDIA}

\date{}
\begin{document}

  \maketitle
\begin{abstract}

We investigate the consequences of generalized CP (GCP) symmetry within the context of the two Higgs doublet model (2HDM), specifically focusing on the lepton sector. Utilizing the Type-I seesaw framework, we study an intriguing connection between the Dirac Yukawa couplings originating from both Higgs fields, leading to a reduction in the number of independent Yukawa couplings and simplifying the scalar and Yukawa sectors when compared to the general 2HDM. The CP3 constraint results in two right-handed neutrinos having equal masses and leads to a diagonal right-handed Majorana neutrino mass matrix. Notably, CP symmetry experiences a soft break due to the phase associated with the vacuum expectation value of the second Higgs doublet. The model aligns well with observed charged lepton masses and neutrino oscillation data, explaining both masses and mixing angles, and yields distinct predictions for normal and inverted neutrino mass hierarchies. It features a novel interplay between atmospheric mixing angle $\theta_{23}$ and neutrino mass hierarchy: the angle $\theta_{23}$ is below maximal for the normal hierarchy and above maximal for inverted hierarchy. Another interesting feature of the model is inherent CP violation for the inverted hierarchy. 

 \end{abstract}

 \section{Introduction}
 
 The standard model (SM) of particle physics provides a unified and well-tested theoretical framework for explaining the interactions of known fundamental particles. It explains how quarks and charged leptons acquire mass. However, it cannot account for the non-zero mass of neutrinos, which is necessary to explain observed neutrino oscillations. One way to naturally explain non-zero neutrino masses is by introducing right-handed neutrinos into the particle content of the SM and allowing them to have a Majorana mass term. This is commonly known as the Type-I seesaw mechanism \cite{Minkowski:1977sc,Yanagida:1979as,Glashow:1979nm,Gell-Mann:1979vob,Mohapatra:1979ia}. The smallness of the neutrino masses can be attained by setting Majorana neutrino mass at high energy scale.

Extending beyond the Standard Model (SM), a natural step involves adding another Higgs doublet, known as the Two-Higgs Doublet Model (2HDM). Initially proposed to address matter-antimatter asymmetry alongside the quark mixing matrix \cite{Lee:1973iz}, the 2HDM doesn't explain neutrino mass. The vacuum expectation values of these two SU(2) doublets, spontaneously break the CP symmetry contributing as an extra source for generating matter-antimatter asymmetry\cite{Lee:1973iz}. Further, the need for a second Higgs doublet arises naturally in the Minimal Supersymmetric Standard Model (MSSM) \cite{Castano:1993ri} and axion models \cite{Weinberg:1977ma,Wilczek:1977pj}. Another reason for considering 2HDM is that it preserves the $\rho$ parameter \cite{Lee:1973iz}, connecting the mechanism of electroweak symmetry breaking with the masses of SM gauge bosons \cite{ParticleDataGroup:2022pth}.

Despite these characteristic features, 2HDMs have shortcomings, including the inability to explain neutrino mass, dark matter, and the allowance of tree-level flavor changing neutral currents (FCNC). The presence of FCNC arises because both SU(2) scalar doublets can couple to fermions. However, there are studies suggesting mechanisms to mitigate FCNC interactions. For example:
\begin{enumerate}
    \item FCNC interactions can be fine-tuned by carefully selecting Yukawa couplings that are suppressed by the heavy mass of the scalar boson responsible for FCNC \cite{Gunion:2002zf}.
    \item FCNC can be eliminated by employing a global symmetry, such as Z${_2}$, which restricts a given scalar boson from coupling to fermions of different electric charges \cite{Glashow:1976nt,Paschos:1976ay}.
    \item Tree-level FCNC can be eliminated by using a global U(1) Peccei-Quinn symmetry \cite{Peccei:1977hh}.
\end{enumerate}
In addition to addressing FCNC, there have been various attempts to enhance 2HDMs to incorporate neutrino masses  \cite{Antusch:2001vn,Atwood:2005bf,Clarke:2015hta,Liu:2016mpf,Camargo:2019ukv,Cogollo:2019mbd} and dark matter \cite{LopezHonorez:2006gr,Gustafsson:2007pc,Dolle:2009fn,LopezHonorez:2010eeh,Arcadi:2018pfo,Bertuzzo:2018ftf}. 
 
However, 2HDM poses a challenge due to its large number of free parameters, making it difficult to probe through collider experiments like the LHC. In general, the scalar potential of the 2HDM consists of fourteen parameters and can exhibit CP conserving or CP violating behavior \cite{Gunion:2002zf,Gunion:2005ja}. Consequently, additional constraints are necessary, often derived from symmetry arguments, to establish relationships among these parameters.

The study of generalized CP (GCP) transformations within the scalar sector of 2HDM is an example of imposing additional symmetries \cite{Ferreira:2009wh,Ferreira:2010bm}. GCP transformations can be categorized in various ways. In Ref.\cite{Ferreira:2009wh}, they are classified into three categories: CP1, CP2, and CP3. CP1 and CP2 correspond to discrete transformations, while CP3 is a continuous transformation that can be extended to the fermionic sector \cite{Ferreira:2010bm}, and they have applied this to the quark sector. Furthermore, the CP symmetries of the scalar sector in the 2HDM have been thoroughly investigated using the basis invariant bilinear formalism \cite{Velhinho:1994np,Ivanov:2006yq,Maniatis:2006fs,Nishi:2006tg,Ivanov:2007de,Maniatis:2007vn} in many works \cite{Ivanov:2006yq,Maniatis:2006fs,Nishi:2006tg,Ivanov:2007de}.   

In 2HDM extensions addressing neutrino mass, explicit neutrino phenomenology, including mixing angles and mass-squared differences, is often absent. In this study, we've extended CP3 to the 2HDM's leptonic Yukawa sector, introducing CP violation through the second Higgs's \textit{vev} phase. Neutrino masses are generated \textit{via} the Type-I seesaw relation, involving right-handed neutrinos. 

The paper is structured as follows: In Section 2, we present the basic formalism of 2HDM. Section 3 elaborates on extending CP3 to the neutrino Yukawa sector within the Type-I seesaw mechanism. We discuss our numerical analysis in Section 4. Finally, in Section 5, we summarize our conclusions.

\section{Two Higgs Doublet Model under Generalized CP Symmetry}

In 2HDMs, the Standard Model's field content expands with the addition of an extra Higgs doublet, denoted as $\Phi_{2}$, which possesses the same charge assignments as the Higgs field in the SM. However, this minimal expansion in the scalar sector results in an increased number of free parameters. To address this parameter growth, it becomes imperative to introduce specific symmetries. In this context, the Generalized CP (GCP) symmetry is considered. Under GCP, scalar doublets undergo transformations as elucidated in \cite{Ferreira:2009wh}:
\begin{equation}
\Phi_a\rightarrow \Phi_{a}^{GCP}=X_{a\alpha}\Phi_{\alpha}^* \,\,,
\end{equation}
where X is an arbitrary unitary CP transformation matrix.

There always exist a choice of basis for which most general GCP transformation matrix can be brought to the form \cite{Ecker:1987qp}
\begin{equation}
X=
\begin{pmatrix}
       \cos \theta && \sin \theta\\
       -\sin \theta && \cos\theta
\end{pmatrix}\,\,,
\end{equation}
where $0 \leq \theta \leq \pi/2$. So, we have three distinct cases with respect to the parameter $\theta$ as mentioned in \cite{Ferreira:2009wh}:
\begin{enumerate}
    \item When $\theta=0$, the symmetry is referred to as CP symmetry of order one (CP1).
    \item For $\theta=\pi/2$, the symmetry is known as CP symmetry of order two (CP2).
    \item In the range $0<\theta<\pi/2$, the symmetry is labeled as CP symmetry of order three (CP3) and importantly it constitutes a continuous symmetry.
\end{enumerate}

The most general scalar potential with two Higgs doublets can be written as
\begin{equation}
\begin{split}
V_H = & \,\, m^2_{11}\Phi_1^{\dagger}\Phi_1+m^2_{22}\Phi_2^{\dagger}\Phi_2-[m^2_{12}\Phi_1^{\dagger}\Phi_2+H.c.]+\frac{1}{2}\lambda_1(\Phi_1^{\dagger}\Phi_1)^2+\frac{1}{2}\lambda_2(\Phi_2^{\dagger}\Phi_2)^2  \\
& +\lambda_3(\Phi_1^{\dagger}\Phi_1)(\Phi_2^{\dagger}\Phi_2)+\lambda_4(\Phi_1^{\dagger}\Phi_2) (\Phi_2^{\dagger}\Phi_1) \\
& +\big[\frac{1}{2}\lambda_5(\Phi_1^{\dagger}\Phi_2)^2 +\lambda_6(\Phi_1^{\dagger}\Phi_1)(\Phi_1^{\dagger}\Phi_2)+\lambda_7(\Phi_2^{\dagger}\Phi_2)(\Phi_1^{\dagger}\Phi_2)+H.c.\big] \,\,,
\end{split}
\end{equation}
 having total 14 parameters. Here, $m_{11}^2$, $m_{22}^2$ and  $\lambda_1$ through $\lambda_4$ are real parameters while $m_{12}^2$, $\lambda_5$, $\lambda_6$ and $\lambda_7$ are generally complex.

Based on the findings in Ref. \cite{Ferreira:2010bm}, it's established that, in addition to the standard CP symmetry CP1, CP3 is the only symmetry that can be extended to the Yukawa sector for leptons. To maintain CP3 invariance within the scalar potential, certain conditions must be satisfied. Specifically, we must have  $m_{11}^2=m_{22}^2$, $m_{12}^2=0$, $\lambda_2=\lambda_1 $, $\lambda_6=0$, $\lambda_7=0$ and  $\lambda_5=\lambda_1-\lambda_3-\lambda_4$ which must be a real parameter.

To avoid the presence of Goldstone bosons following spontaneous symmetry breaking, it's necessary to introduce soft CP3 symmetry breaking. Therefore, we will consider $m_{11}^2\neq m_{22}^2$ and $\Re[m_{12}^2]\neq 0$. This softly broken CP3 symmetric potential also leads to a CP-violating vacuum expectation value (\textit{vev}) for the second Higgs doublet, which assumes a crucial role in the exploration of CP violation in the lepton sector, as we will delve into in the upcoming sections.

\section{CP3 in Yukawa Sector with Type-I seesaw}

In our study, we've expanded upon the Standard Model (SM) by incorporating one Higgs doublet and three right-handed neutrinos denoted as $N_R$. Within the framework of the Type-I seesaw mechanism, the relevant Yukawa Lagrangian responsible for generating the masses of both charged leptons and neutrinos\footnote{For quark masses, see Ref. \cite{Ferreira:2010bm}} is expressed as follows 
\begin{equation}\label{lag}
-\mathcal{L}_Y=\overline{L}_L \Gamma_a\Phi_a l_R+\overline{L}_L Y_a\tilde{\Phi}_a N_R+\frac{1}{2}\overline{N^c_R}M N_R +H.c.\,\,,
\end{equation}
where $L_L$, $l_R$ are Standard Model $SU(2)$ left-handed doublets and right-handed singlets, $\Phi_a $ (a = 1, 2) are Higgs doublets and N$_R$ are right-handed neutrino singlets. $\Gamma_a$ and $Y_a$ are the Yukawa coupling matrices for charged leptons and neutrinos respectively and M is lepton number violating Majorana mass term for right-handed neutrinos. Now we will extend the GCP symmetry to the leptonic Yukawa sector.

The fields involved in Eqn.(\ref{lag}) transforms under GCP symmetry as
\begin{equation}
\begin{rcases}
\begin{aligned}
    \Phi_a &\rightarrow \Phi_{a}^{GCP}=X_{ab}\Phi_{b}^*\,\,,\\
    \tilde{\Phi}_a &\rightarrow \tilde{\Phi}_{a}^{GCP}=  X_{ab}^*(\tilde{\Phi}_{b}^\dagger)^T\,\,,\\
   L_L &\rightarrow L_{L}^{GCP}= i X_{\zeta}\gamma^0C\overline{L}_{L}^T\,\,,\\
   l_R &\rightarrow l_{R}^{GCP}= i X_{\beta}\gamma^0C\overline{l}_{R}^T\,\,,\\
   N_R &\rightarrow N_{R}^{GCP}= i X_{\gamma}\gamma^0C\overline{N}_{R}^T\,\,,
\end{aligned}
\label{eqgroup}
\end{rcases}
\end{equation}
where $\gamma^0$ is Dirac gamma matrix and C is charge conjugation matrix, $X$, $ X_{\zeta}$, $X_{\beta}$ and $X_{\gamma}$ are CP transformation matrices.

For Lagrangian to remain invariant under these CP transformations we find  Yukawa coupling matrices to transform as
\begin{equation}\label{gama1}
\Gamma_b^*=X_{\zeta}^{\dagger}\Gamma_a X_{\beta} X_{ab}\,\,,
\end{equation}
\begin{equation}\label{y1}
Y_b^*=X_{\zeta}^{\dagger}Y_a X_{\gamma} X_{ab}^*\,\,,
\end{equation}
and Majorana mass matrix  to transform as
\begin{equation}\label{M1}
M^*=X_{\gamma}^{T}M X_{\gamma}\,\,,
\end{equation}
where
\begin{equation}\label{Mmatrix}
M=
\begin{pmatrix}
       M_{11} &&  M_{12} &&  M_{13}\\
      M_{12} &&  M_{22} &&  M_{23}\\
        M_{13} &&  M_{23}&&  M_{33}
\end{pmatrix}\,\,.
\end{equation}
The CP  transformation matrices involved are given by
\begin{equation}
X_{\zeta}=
\begin{pmatrix}
       \cos\zeta&& \sin\zeta&& 0\\
       -\sin\zeta && \cos\zeta &&0\\
       0&&0&&1
\end{pmatrix}\,\,,
\end{equation}
\begin{equation}
X_{\beta}=
\begin{pmatrix}
       \cos\beta&& \sin\beta&& 0\\
       -\sin\beta && \cos\beta &&0\\
       0&&0&&1
\end{pmatrix}\,\,,
\end{equation}
\begin{equation}\label{Xg}
X_{\gamma}=
\begin{pmatrix}
       \cos\gamma&& \sin\gamma&& 0\\
       -\sin\gamma && \cos\gamma &&0\\
       0&&0&&1
\end{pmatrix}\,\,.
\end{equation}

It was found in Ref. \cite{Ferreira:2010bm} that  CP3 symmetry with $\theta=\pi/3$ ($\zeta=\beta=\gamma=\pi/3$) can be extended to Yukawa sector producing correct quark masses. Under these conditions, the forms of Yukawa coupling matrices given in Eqns.(\ref{gama1}) and (\ref{y1}) become
\begin{equation}
\Gamma_1=
\begin{pmatrix}\label{gs}
       i a_{11}&&i a_{12}&& a_{13}\\
      i a_{12}&& -i a_{11} &&a_{23}\\
       a_{31}&&a_{32}&&0
\end{pmatrix}
\,\,,\,\,\
\Gamma_2=
\begin{pmatrix}
        i a_{12}&&-ia_{11}&& -a_{23}\\
      -i a_{11}&& -i a_{12} &&a_{13}\\
       -a_{32}&&a_{31}&&0
\end{pmatrix}\,\,,
\end{equation}
\begin{equation}
Y_1=
\begin{pmatrix}\label{ys}
       i b_{11}&&i b_{12}&& b_{13}\\
      ib_{12}&& -i b_{11} &&b_{23}\\
       b_{31}&&b_{32}&&0
\end{pmatrix}
\,\,,\,\,\
Y_2=
\begin{pmatrix}
        i b_{12}&&-i b_{11}&& -b_{23}\\
      -i b_{11}&& -ib_{12} &&b_{13}\\
       -b_{32}&&b_{31}&&0
\end{pmatrix}\,\,,
\end{equation}
where all $a$'s and $b$'s are real parameters. The choice of $\theta=\pi/3$ alongside $\zeta=\beta=\gamma=\pi/3$ in the leptonic sector stems from the similarity in GCP transformation properties between quarks and leptonic fields, as outlined in Eqn.(\ref{eqgroup}).

Now, we need to solve constrains given by Eqn.(\ref{M1}) which can be  rewritten as:
\begin{equation}\label{M2}
M^*-X_{\gamma}^{T}M X_{\gamma}=0\,\,.
\end{equation}
Using Eqns.(\ref{Mmatrix}) and (\ref{Xg}) in  Eqn.(\ref{M2}), the set of constraints are
\begin{eqnarray}
   M_{11}^*-M_{11} \cos^2 \gamma -M_{22} \sin^2\gamma+M_{12} \sin2 \gamma&=&0,\label{M11}\\
    M_{12}^*-M_{12} \cos2\gamma +(-M_{11} + M_{22}) \sin\gamma \cos\gamma&=&0, \label{M12}\\ 
   M_{22}^*-M_{22} \cos^2\gamma -2 M_{12} \cos\gamma\sin\gamma-M_{11} \sin^2\gamma&=&0, \label{M22}\\ 
   M_{13}^*-M_{13} \cos\gamma+M_{23} \sin\gamma& =&0,\label{M13}\\ 
   M_{23}^*-M_{23} \cos\gamma-M_{13} \sin\gamma&=&0,\label{M23}\\  
   M_{33}^*-M_{33} &=&0. \label{M33}
\end{eqnarray}
In Eqns.(\ref{M11}), (\ref{M12}) and (\ref{M22}), the real part can be separated out as
\begin{equation}
\begin{pmatrix}
1-\cos^2\gamma & \sin2 \gamma  & -\sin^2\gamma \\
-\cos\gamma \sin\gamma &  1-\cos 2\gamma & \cos\gamma \sin\gamma \\
-\sin^2\gamma& -2 \cos\gamma \sin\gamma & 1-\cos^2\gamma\\
\end{pmatrix}
\begin{pmatrix}
\Re[M_{11}]\\
\Re[M_{12}] \\
\Re[M_{22}]\\
\end{pmatrix}
=0,
\end{equation}
and the imaginary part can be separated out as
\begin{equation}
\begin{pmatrix}
-1-\cos^2\gamma & \sin2 \gamma  & -\sin^2\gamma \\
-\cos\gamma \sin\gamma & - 1-\cos 2\gamma & \cos\gamma \sin\gamma \\
-\sin^2\gamma& -2 \cos\gamma \sin\gamma & -1-\cos^2\gamma\\
\end{pmatrix}
\begin{pmatrix}
\Im[M_{11}]\\
\Im[M_{12}] \\
\Im[M_{22}]\\
\end{pmatrix}
=0.
\end{equation}.
Further, from Eqns.(\ref{M13}) and (\ref{M23}) we have, for real part
\begin{equation}\label{Re1323}
\begin{pmatrix}
1-\cos\gamma & \sin\gamma \\
-\sin\gamma & 1-\cos\gamma  \\
\end{pmatrix}
\begin{pmatrix}
\Re[M_{13}]\\
\Re[M_{23}]
\end{pmatrix}
=0,
\end{equation}
and, for imaginary part,
\begin{equation}\label{Im1323}
\begin{pmatrix}
-1-\cos\gamma & \sin\gamma \\
-\sin\gamma & -1-\cos\gamma  \\
\end{pmatrix}
\begin{pmatrix}
\Im[M_{13}]\\
\Im[M_{23}]
\end{pmatrix}
=0.
\end{equation}
For  $\gamma = \pi/3$, we have
\begin{equation}\label{Re}
\begin{pmatrix}
\frac{3}{4}& \frac{\sqrt{3}}{2} &\frac{-3}{4} \\
-\frac{\sqrt{3}}{4} & \frac{3}{2} & \frac{\sqrt{3}}{4} \\
\frac{-3}{4}& \frac{-\sqrt{3}}{2} & \frac{3}{4}
\end{pmatrix}
\begin{pmatrix}
\Re[M_{11}]\\
\Re[M_{12}] \\
\Re[M_{22}]\\
\end{pmatrix}
=0,
\end{equation}
\begin{equation}\label{Im}
\begin{pmatrix}
\frac{-5}{4}& \frac{\sqrt{3}}{2} &\frac{-3}{4} \\
-\frac{\sqrt{3}}{4} & -\frac{1}{2} & \frac{\sqrt{3}}{4} \\
\frac{-3}{4}& -\frac{\sqrt{3}}{2} & -\frac{5}{4}
\end{pmatrix}
\begin{pmatrix}
\Im[M_{11}]\\
\Im[M_{12}] \\
\Im[M_{22}]\\
\end{pmatrix}
=0,
\end{equation}
and
\begin{equation}\label{RE1323}
\begin{pmatrix}
\frac{1}{2} &\frac{\sqrt{3}}{2} \\
-\frac{\sqrt{3}}{2}  & \frac{1}{2}  \\
\end{pmatrix}
\begin{pmatrix}
\Re[M_{13}]\\
\Re[M_{23}]
\end{pmatrix}
=0,
\end{equation}
\begin{equation}\label{IM1323}
\begin{pmatrix}
-\frac{3}{2} &\frac{\sqrt{3}}{2} \\
-\frac{\sqrt{3}}{2}  & -\frac{3}{2}  \\
\end{pmatrix}
\begin{pmatrix}
\Im[M_{13}]\\
\Im[M_{23}]
\end{pmatrix}
=0.
\end{equation}

In Eqns.(\ref{Im}), (\ref{RE1323}), and (\ref{IM1323}), the determinant of the square matrix is non-zero, implying a unique solution where $\Im[M_{11}]$, $\Im[M_{12}]$, $\Im[M_{22}]$, $\Re[M_{13}]$, $\Re[M_{23}]$, $\Im[M_{13}]$, and $\Im[M_{23}]$ all equal zero. On the other hand, in Eqn.(\ref{Re}), the determinant of the square matrix is zero, indicating arbitrary solutions, with $\Re[M_{12}]$ equal to zero and $\Re[M_{11}]\equiv M_1$ equal to $\Re[M_{22}]$. Furthermore, Eqn.(\ref{M33}) leads to $\Im[M_{33}]$ being zero, leaving only $\Re[M_{33}]\equiv M_3$ as the relevant parameter.

It's worth noting that the CP3 constraint results in two right-handed neutrinos having equal masses, leading to a diagonal matrix $M$, described as follows
\begin{equation}\label{mr}
M=
\begin{pmatrix}
       M_1&&0&& 0\\
      0&&M_1 &&0\\
       0&&0&&M_3
\end{pmatrix}\,\,.
\end{equation}
After spontaneous symmetry breaking (SSB), both Higgs doublets get  \textit{vevs}, given by
\begin{equation}
\left<\Phi_1\right>=
\begin{pmatrix}
       0\\
      \frac{v_1}{\sqrt{2}}
\end{pmatrix}
 \,\,\,,\,\,
\left<\Phi_2\right>=
\begin{pmatrix}
       0\\
      e^{i\alpha}\frac{v_2 }{\sqrt{2}}
\end{pmatrix}\,\,,
\end{equation}
with condition that $v=\sqrt{v_1^2+v_2^2}\approx 245 $ GeV, where $v$ is the standard model \textit{vev}. Consequently, the charged leptons mass matrix becomes
\begin{eqnarray}
 M_l&=&\frac{1}{\sqrt{2}}(v_1\Gamma_1+e^{i\alpha} v_2 \Gamma_2),\\
 &=&\frac{1}{\sqrt{2}}(\cos\phi\Gamma_1+e^{i\alpha} \sin\phi \Gamma_2) v,
\end{eqnarray}
and for neutrinos we have, Dirac mass matrix given by
\begin{eqnarray}
 M_D&=&\frac{1}{\sqrt{2}}(v_1 Y_1+e^{-i\alpha} v_2 Y_2),\\
&=&\frac{1}{\sqrt{2}}(\cos\phi Y_1+e^{-i\alpha} \sin\phi Y_2) v,\label{md2}
\end{eqnarray}
where $\phi$ is defined as $\tan\phi=v_2/v_1$. 

We work in the basis in which charged lepton mass matrix is diagonal. The charged lepton mass matrix can be diagonalized as
\begin{equation}
    M_l=U_l m_{diag}U_R^{\dagger},
\end{equation}
where $U_l$ and $U_R$ are $3\times3$ unitary matrices and $m_{diag}=diag(m_e, m_{\mu}, m_{\tau})$ is diagonal matrix with positive real entries giving mass eigenvalues of electron, muon, and tau, respectively. So, we have
\begin{eqnarray}\label{mld}
    U_l^{\dagger}M_lM_l^{\dagger}U_l=m_{diag}^2,
\end{eqnarray}
where $U_l$ rotates $M_D$ into the basis in which charged leptons are diagonal
\begin{equation}
   M_D^{new}=U_l^{\dagger}M_D. 
\end{equation}
Using Type-I seesaw, the effective light neutrinos mass matrix is given by
\begin{eqnarray}
M_{\nu}&=&-M_D^{new} M^{-1} (M_D^{new})^T\,,\\
&=&-(U_l^{\dagger}M_D ) M^{-1} (U_l^{\dagger}M_D)^T,\label{mnuf}
\end{eqnarray}
which is a complex symmetric matrix. This matrix is related to Yukawa coupling matrices $Y_1$ and $Y_2$ through Eqn.(\ref{md2}). The effective light neutrino mass matrix can be diagonalized by $3\times3$ unitary matrix $U$ as
\begin{equation}\label{mdd}
    U^{\dagger}M_{\nu}U^*=m,
\end{equation} where $m_{ik}=m_i \delta_{ik}$, $m_i>0$
$(i,k=1, 2, 3)$.

We will now move forward with the numerical determination of charged lepton masses and the parameters governing neutrino oscillations. This process entails the variation of free parameters to ascertain the permissible parameter space within the model.

\section{Numerical Analysis and Discussion}

In our numerical analysis, we generated random numbers uniformly for the \textit{vev}-phase $\alpha$ in the range of 0 to 2$\pi$. We also generated random numbers uniformly for the masses of the right-handed neutrinos $M_1$ and $M_3$, which ranged from 10$^{11}$ to 10$^{13}$ GeV and from 1.1$\times$10$^{13}$ to 10$^{15}$ GeV, respectively. We considered two cases for the \textit{vev} $v_1$, as discussed in the following subsections.

\subsection{When $v_1<<v_2$}

In this scenario, we examined the influence of a very small \textit{vev} on our parameter space. To emphasize the dominance of \textit{vev} $v_2$, we randomly varied $v_1$ in the range of $(0-5)\times10^{-6}$ GeV. The value of $v_2$ is subsequently determined using the equation $v_2=\sqrt{v^2-v_1^2}$ GeV. We then determined the masses of charged leptons and the parameters governing neutrino oscillations, which we discuss in the following subsections.

\subsubsection{Charged Lepton Masses}

To compute the masses of charged leptons, we varied the Yukawa coupling parameters within the range specified in Table \ref{data}. We then proceeded to numerically diagonalize the mass matrix $M_lM_l^{\dagger}$, as described in Eqn.(\ref{mld}), to obtain the squared masses of charged leptons ($m_e^2,m_\mu^2,m_\tau^2$). Through this analysis, we identified parameter values that consistently yielded the correct charged lepton masses for both normal ($m_1<m_2<m_3$) and inverted ($m_3<m_1<m_2$) hierarchies of neutrinos. The benchmark points are listed in the second column of Table \ref{data3}. With these parameter values, we calculated the charged lepton masses, as presented in Table \ref{data5}, which closely align with experimentally observed values.

\begin{table}[t]\label{data}
\begin{center}
\begin{tabular}{|l|l|l|}
\hline\hline
Yukawa Coupling  & \hspace{0.2cm} When $v_1<<v_2$  & When $v_1$ ranges from  ($10-17$) Gev  \\
\hspace{0.7cm}Parameter & \hspace{0.9cm} Range & \hspace{2.7cm}Range\\
\hline
\hspace{0.7cm}$a_{11}$ & $3\times 10^{-5} - 6\times 10^{-5}$& \hspace{1.7cm}$2\times 10^{-5} - 3\times 10^{-5}$\\
\hspace{0.7cm}$a_{12}$ & $8\times 10^{-5} - 1\times 10^{-4}$& \hspace{1.7cm}$6\times 10^{-5} - 8\times 10^{-5}$\\
\hspace{0.7cm}$a_{13}$ & $6\times 10^{-3} - 8\times 10^{-3}$& \hspace{1.7cm}$6\times 10^{-3} - 7\times 10^{-3}$\\
\hspace{0.7cm}$a_{23}$ & $6\times 10^{-3} - 8\times 10^{-3}$&\hspace{1.7cm} $8\times 10^{-3} - 9\times 10^{-3}$ \\
\hspace{0.7cm}$a_{31}$ & $4\times 10^{-4} - 6\times 10^{-4}$& \hspace{1.7cm}$4\times 10^{-4} - 6\times 10^{-4}$\\
\hspace{0.7cm}$a_{32}$ & $1\times 10^{-4} - 2\times 10^{-4}$& \hspace{1.7cm}$1\times 10^{-4} - 2\times 10^{-4}$\\
\hline\hline
\end{tabular}
\end{center}
\caption{\label{data}The ranges of Yukawa couplings used in  numerical analysis for charged leptons.}
\end{table}

\begin{figure}[t]
  \centering
\begin{tabular}{cc}
\includegraphics[width=0.45\linewidth]{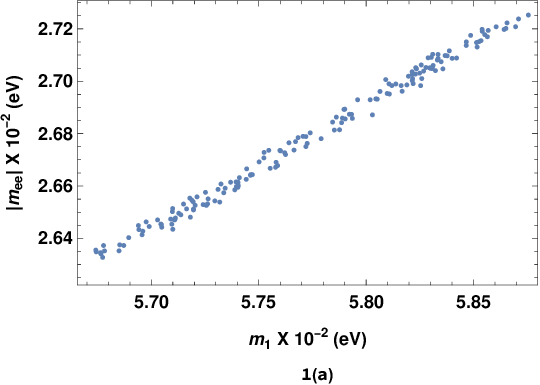}& \includegraphics[width=0.45\linewidth]{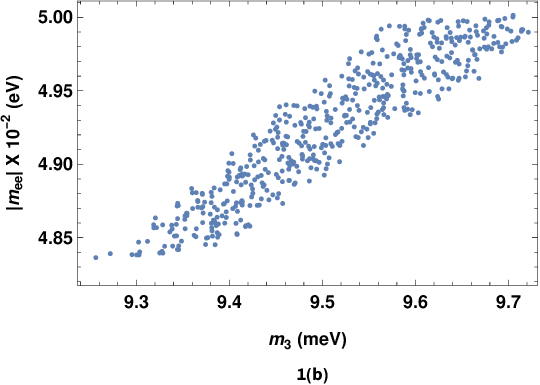}\\
\end{tabular}
  \caption{Predictions for effective Majorana mass $|m_{ee}| $  for normal (left) and inverted (right) hierarchy  when $v_1<<v_2$.}
  \label{fig1}
\end{figure}

\begin{figure}[t]
  \centering
\begin{tabular}{cc}
\includegraphics[width=0.45\linewidth]{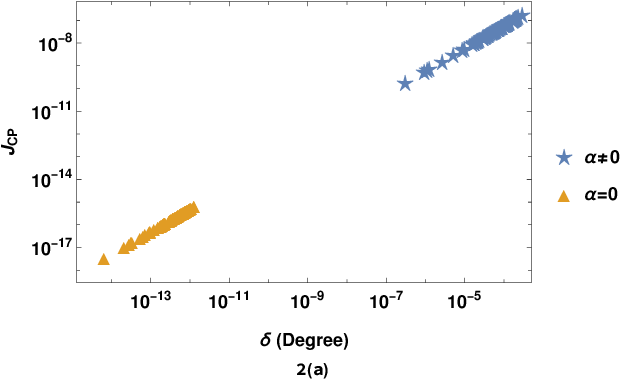}& \includegraphics[width=0.45\linewidth]{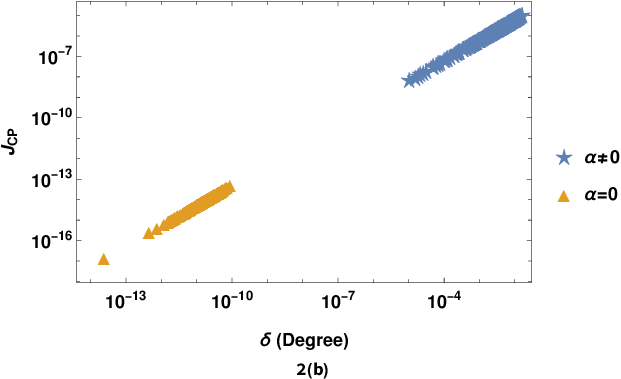}\\
\end{tabular}
  \caption{$\delta-J_{CP}$ correlation with \textit{vev}-phase $\alpha=0$ and $\alpha\neq 0$ for normal (left) and inverted (right) hierarchies.}
  \label{fig2}
\end{figure}

\begin{figure}[t]
  \centering
\begin{tabular}{cc}
\includegraphics[width=0.45\linewidth]{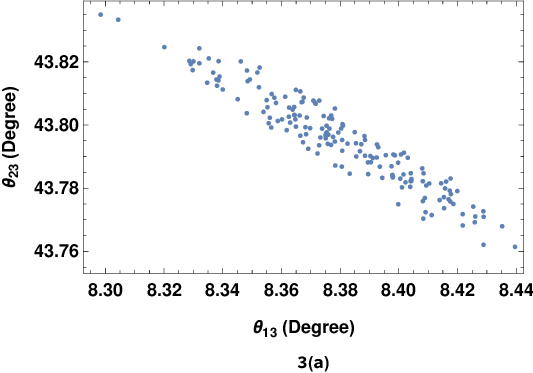}& \includegraphics[width=0.45\linewidth]{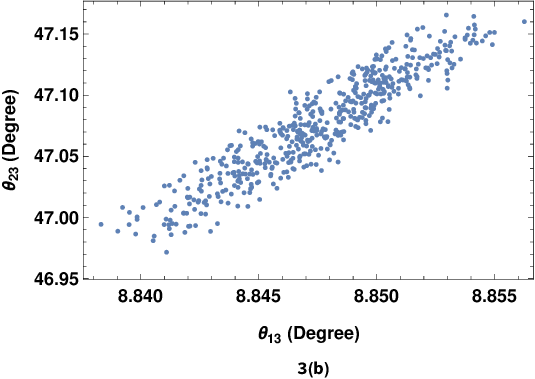}\\
\end{tabular}
  \caption{Correlation between the mixing angles $\theta_{13}$ and $\theta_{23}$ for normal (left) and inverted (right) hierarchies.}
  \label{fig3}
  \end{figure}

\begin{figure}[t]
  \centering
\begin{tabular}{cc}
\includegraphics[width=0.45\linewidth]{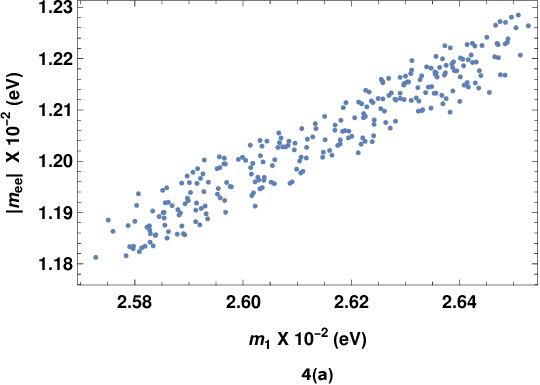}& \includegraphics[width=0.45\linewidth]{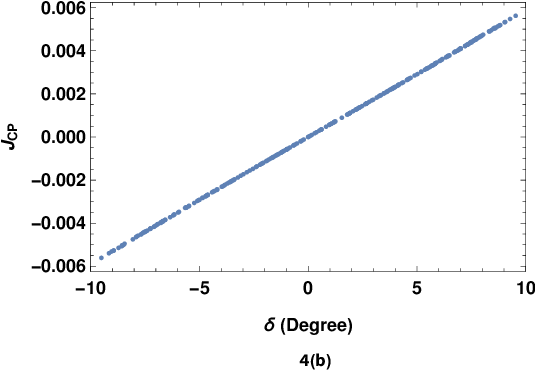}\\
 \includegraphics[width=0.45\linewidth]{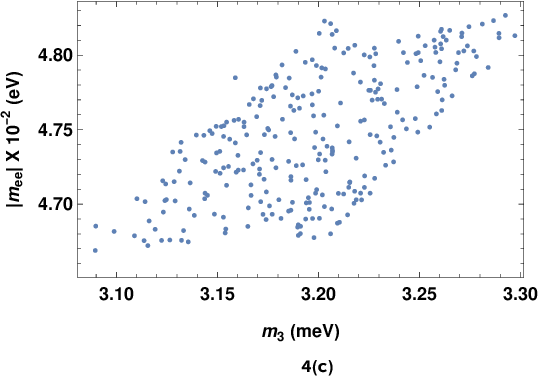} & \includegraphics[width=0.45\linewidth]{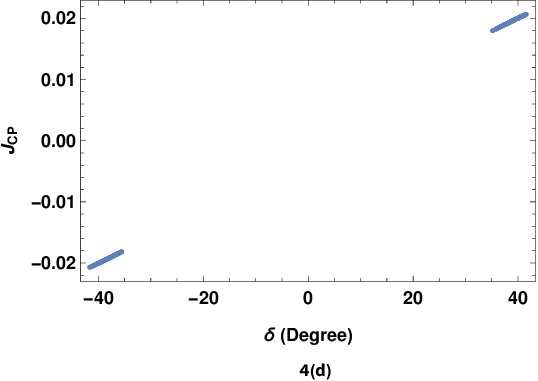}

\end{tabular}
  \caption{Correlation plots between the effective neutrino mass $|m_{ee}|$ and lightest neutrino mass, as well as between $J_{CP}$ and CP-violating phase $\delta$, for both normal (first row) and inverted (second row) hierarchies when the \textit{vev} $v_1$ ranges from 10 to 17 GeV.}
  \label{fig5}
\end{figure}
\begin{figure}[t]
  \centering
\begin{tabular}{cc}
\includegraphics[width=0.4\linewidth]{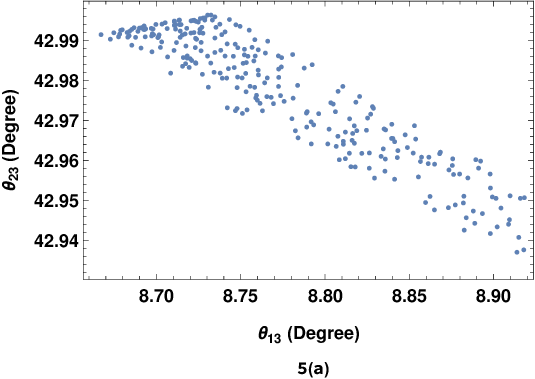}& \includegraphics[width=0.4\linewidth]{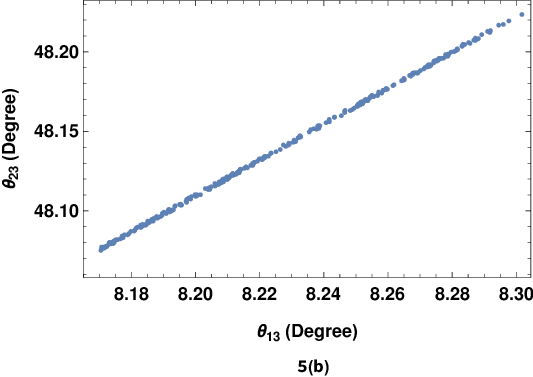} 
\end{tabular}
  \caption{Correlation between the mixing angles $\theta_{13}$ and $\theta_{23}$ for normal (left) and inverted (right) hierarchies.}
  \label{fig6}
\end{figure}

\begin{figure}[t]
  \centering
\begin{tabular}{cc}
\includegraphics[width=0.4\linewidth]{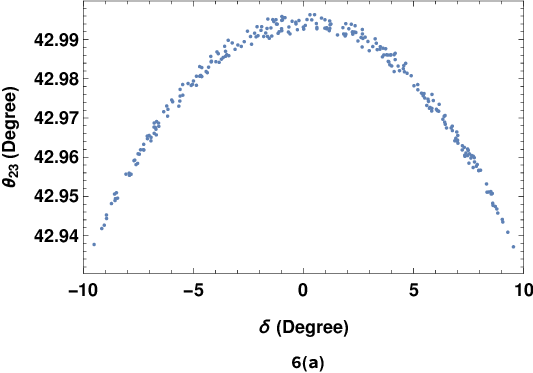}& \includegraphics[width=0.4\linewidth]{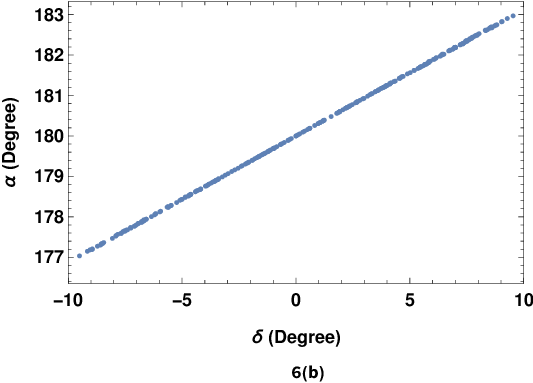} \\
    
    \includegraphics[width=0.4\linewidth]{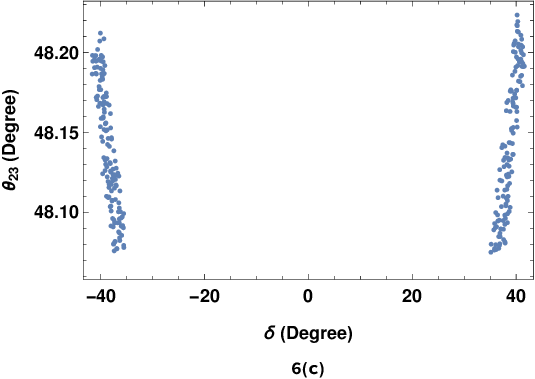}& \includegraphics[width=0.4\linewidth]{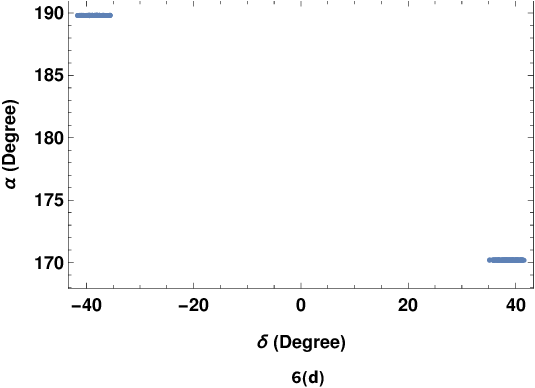}
\end{tabular}
  \caption{Correlation of $\delta$ with $\theta_{23}$ and \textit{vev}-phase $\alpha$ for NH (first row) and IH (second row).}
  \label{fig7}
\end{figure}


\begin{table}\label{data3}

\begin{center}
\begin{small}
\begin{tabular}{|l|l|l|l|}
\hline\hline
Parameters  & When $v_1<<v_2$ &\multicolumn{2}{c|}{When $v_1$ ranges from  ($10-17$) GeV }  \\
\cline{3-4}
 (Units)& & NH&IH\\
\hline
\hspace{0.5cm}$a_{11}$ &  $5.740882\times 10^{-5}$& $2.692411\times 10^{-5}$&$2.692412\times 10^{-5}$\\
\hspace{0.5cm}$a_{12}$ & $9.066478\times 10^{-5}$  & $7.993756\times 10^{-5}$&$7.993757\times10^{-5}$\\
\hspace{0.5cm}$a_{13}$ & $7.124930\times 10^{-3}$ & $6.382725\times 10^{-3}$&$6.382726\times10^{-3}$\\
\hspace{0.5cm}$a_{23}$ & $7.377109\times 10^{-3}$& $8.028028\times 10^{-3}$&$8.028028\times10^{-3}$\\
\hspace{0.5cm}$a_{31}$ & $5.843790\times 10^{-4}$& $5.945601\times 10^{-4}$&$5.945602\times10^{-4}$\\
\hspace{0.5cm}$a_{32}$ &$1.377038\times 10^{-4}$ & $1.065540\times 10^{-4}$&$1.065540\times10^{-4}$\\
\hspace{0.5cm}$v_1$ (GeV) &$3.287415\times 10^{-6}$& $11.00$&$15.30$\\
\hspace{0.5cm}$v_{2}$ (GeV) &$245$  & $244.75$&$244.52$\\
\hspace{0.5cm}$\alpha$ ($^o$) &$197.54$& $177.68$&$170.20$\\
\hline\hline
\end{tabular}
\end{small}
\end{center}
\caption{\label{data3}Benchmark points for both the cases (i) $v_1<<v_2$ (second column) (ii) $v_1$ ranges from  ($10-17$) GeV (third column), yielding correct values of charged lepton masses.}
\end{table}  


\begin{table}\label{data5}

\begin{center}
\begin{small}
\begin{tabular}{|l|l|l|l|}
\hline\hline
Masses & When $v_1<<v_2$ &  \multicolumn{2}{c|}{When $v_1$ ranges from  ($10-17$) Gev } \\
 \cline{3-4}
(MeV)& &NH&IH \\
\hline
\hspace{0.5cm}$m_{e}$ &  $0.510$&  $0.511$&$0.511$\\
\hspace{0.5cm}$m_{\mu}$ & $105.66$ & $105.66$&$105.66$\\
\hspace{0.5cm}$m_{\tau}$ & $1776.87$ &  $1776.85$&$1776.85$\\
\hline\hline
\end{tabular}
\end{small}
\end{center}
\caption{\label{data5} The values of the charged lepton masses obtained using benchmark points given in Table \ref{data3}.}
\end{table}   

\begin{table}\label{data2}
\begin{center}
\begin{small}
\begin{tabular}{|l|l|l|l|l|}
\hline\hline
Yukawa   & \multicolumn{2}{c|}{When $v_1<<v_2$} & \multicolumn{2}{c|}{When $v_1$ ranges from  ($10-17$) Gev}  \\
Coupling& \multicolumn{2}{c|}{Range}& \multicolumn{2}{c|}{Range} \\
\cline{2-5}
Parameter &\hspace{1.3cm}NH&\hspace{1.3cm}IH&\hspace{1.3cm}NH&\hspace{1.3cm}IH\\
\hline
\hspace{0.5cm}$b_{11}$ &  $6\times 10^{-2}-8\times 10^{-2}$&  $7\times 10^{-2} - 2\times 10^{-1}$ &  $6\times 10^{-2} - 8\times 10^{-2}$& $8\times 10^{-2} -9\times 10^{-2}$\\
\hspace{0.5cm}$b_{12}$ & $1\times 10^{-2} - 5\times 10^{-2}$ & $2\times 10^{-3} - 4\times 10^{-3}$& $2\times 10^{-2} - 3\times 10^{-2}$ & $2\times 10^{-3} - 3\times 10^{-3}$\\
\hspace{0.5cm}$b_{13}$ & $7\times 10^{-1} - 9\times 10^{-1}$ & $3\times 10^{-1} - 5\times 10^{-1}$& $6\times 10^{-1} - 8\times 10^{-1}$& $1\times 10^{-1} - 2\times 10^{-1}$\\
\hspace{0.5cm}$b_{23}$ & $1\times 10^{-4} - 5\times 10^{-4}$& $4\times 10^{-1} - 6\times 10^{-1}$& $1\times 10^{-3} - 2\times 10^{-3}$& $4\times 10^{-1} - 6\times 10^{-1}$\\
\hspace{0.5cm}$b_{31}$ & $1\times 10^{-4} - 5\times 10^{-4}$& $3\times 10^{-2} - 5\times 10^{-2}$& $1\times 10^{-2} -2\times 10^{-2}$& $3\times 10^{-2} - 5\times 10^{-2}$\\
\hspace{0.5cm}$b_{32}$ &$4\times 10^{-2} - 6\times 10^{-2}$& $3\times 10^{-2} - 5\times 10^{-2}$& $4\times 10^{-2} -6\times 10^{-2}$ & $6\times 10^{-2} - 7\times 10^{-2}$\\
\hline\hline
\end{tabular}
\end{small}
\end{center}
\caption{\label{data2} The ranges of Yukawa couplings used in numerical analysis for normal and inverted hierarchies of neutrinos.}
\end{table}

\begin{table}\label{data4}

\begin{center}
\begin{small}
\begin{tabular}{|l|l|l|l|l|}
\hline\hline
Parameter & \multicolumn{2}{c|}{When $v_1<<v_2$} & \multicolumn{2}{c|}{When $v_1$ ranges from  ($10-17$) Gev}  \\
 \cline{2-5}
(Units) &\hspace{1.3cm}NH&\hspace{1.3cm}IH&\hspace{1.3cm}NH&\hspace{1.3cm}IH\\
\hline
\hspace{0.5cm}$b_{11}$ &  $7.244577\times 10^{-2}$&  $1.066963\times 10^{-1}$ &  $6.009812\times 10^{-2}$& $8.006565\times 10^{-2}$\\
\hspace{0.5cm}$b_{12}$ & $1.498695\times 10^{-2}$ & $3.644816\times 10^{-3}$& $2.334813\times 10^{-2}$ & $2.601354\times 10^{-3}$\\
\hspace{0.5cm}$b_{13}$ & $8.560231\times 10^{-1}$ & $4.519143\times 10^{-1}$& $6.935666\times 10^{-1}$& $1.522807\times 10^{-1}$\\
\hspace{0.5cm}$b_{23}$ & $1.033423\times 10^{-4}$& $5.112375\times 10^{-1}$& $1.791234\times 10^{-3}$& $4.531115\times 10^{-1}$\\
\hspace{0.5cm}$b_{31}$ & $1.356340\times 10^{-4}$& $4.546001\times 10^{-2}$& $1.655557\times 10^{-2}$& $3.909223\times 10^{-2}$\\
\hspace{0.5cm}$b_{32}$ &$5.796197\times 10^{-2}$& $4.435139\times 10^{-2}$& $4.880238\times 10^{-2}$ & $6.157254\times 10^{-2}$\\
\hspace{0.5cm}$M_1$ (GeV) &$2.447917\times 10^{12}$& $6.721219\times 10^{12}$& $3.066113\times 10^{12}$ & $4.417490\times 10^{12}$\\
\hspace{0.5cm}$M_{3}$ (GeV) &$2.222544\times 10^{14}$ & $5.140671\times 10^{14}$& $2.748449\times 10^{14}$& $6.211162\times 10^{14}$\\
\hline\hline
\end{tabular}
\end{small}
\end{center}
\caption{\label{data4}The benchmark points yielding correct neutrino phenomenology (i.e. neutrino mixing angles and mass-squared differences are within $3\sigma$ experimental range\cite{deSalas:2020pgw}) for normal and inverted hierarchical neutrino masses.}
\end{table}  

\begin{table}\label{data6}

\begin{center}
\begin{small}
\begin{tabular}{|l|l|l|l|l|}
\hline\hline
Parameters & \multicolumn{2}{c|}{When $v_1<<v_2$} & \multicolumn{2}{c|}{When $v_1$ ranges from  ($10-17$) Gev}  \\
 \cline{2-5}
 (Units)&\hspace{1.3cm}NH&\hspace{1.3cm}IH&\hspace{1.3cm}NH&\hspace{1.3cm}IH\\
\hline
\hspace{0.5cm}$\theta_{13}$ ($^o$) &  $8.44$&  $8.85$ &  $8.86$& $8.22$\\
\hspace{0.5cm}$\theta_{12}$ ($^o$)& $31.70$ & $34.30$& $32.10$ & $31.80$\\
\hspace{0.5cm}$\theta_{23}$ ($^o$)& $43.76$ & $47.09$& $42.96$& $48.14$\\
\hspace{0.5cm}$\Delta m_{21}^2$ (eV$^2$)& $7.61\times10^{-5}$& $7.33\times10^{-5}$& $7.64\times10^{-5}$& $7.04\times10^{-5}$\\
\hspace{0.5cm}$\Delta m_{31}^2$ (eV$^2$) & $2.58\times10^{-3}$& $2.48\times10^{-3}$& $2.49\times10^{-3}$& $2.43\times10^{-3}$\\
\hline\hline
\end{tabular}
\end{small}
\end{center}
\caption{\label{data6}The values of the neutrino oscillation parameters obtained using benchmark points given in Table \ref{data4} for normal and inverted hierarchical neutrino masses.}
\end{table} 

\subsubsection{Neutrino Oscillation Parameters}

To find the neutrino mixing angles and mass-squared differences, we first rotated the neutrino mass matrix using the previously derived $U_l$. The matrix $M_\nu$, described in Eqn.(\ref{mnuf}), generally contains Yukawa couplings $Y_1$ and $Y_2$, as given in Eqn.(\ref{ys}). We then introduced random variations to these Yukawa coupling parameters within the specified ranges presented in the second column of Table \ref{data2}, considering both normal (NH) and inverted (IH) hierarchies. This process allowed us to identify a benchmark point that consistently yielded values for the neutrino oscillation parameters\cite{deSalas:2020pgw}. These values are listed in the second column of Table \ref{data4} for reference. The corresponding values of the mixing angles and mass-squared differences can be found in the second column of Table \ref{data6}.
 
For these parameter values, the effective Majorana neutrino mass takes the values $|m_{ee}|=\left|\sum_i U_{ei}^2m_i\right|=0.02693$ eV and $|m_{ee}|=0.04952$ eV for NH and IH of neutrinos, respectively.  The masses $m_1$, $m_2$, and $m_3$ exhibited significant degeneracy in the case of NH, while for IH, there was an order of magnitude difference between the lightest mass and the other heavier masses. A linear correlation between $|m_{ee}|$ and the lightest neutrino mass ($m_1$ for NH and $m_3$ for IH) is evident in Fig. \ref{fig1}.

CP violation in the leptonic sector remains unobserved to date. Consequently, it will be instructive to scrutinize the model's predictions for CP violation. In our model, CP violation, effectively, seeds from the complex \textit{vev} $\left<\Phi_2\right>$ \textit{via} its phase $\alpha$. CP violation can be understood in a rephasing invariant way by defining Jarlskog parameter $J_{CP}=\Im\left(U_{\mu 3}U_{e3}^* U_{e2}U_{\mu 2}^*\right)$ \cite{Jarlskog:1985ht,Bilenky:1987ty,Krastev:1988yu}. For the present case, i.e. $v_1<<v_2$, we find that $J_{CP}$ and $\delta$ (the phase of the $U_{e3}$ element in $U$) are exceedingly small, regardless of the neutrino mass hierarchy, and whether the \textit{vev}-phase $\alpha$ is equal to zero or non-zero (see Fig. (\ref{fig2})). Figure (\ref{fig2}) clearly illustrates that when $v_1<<v_2$, the model predicts a effective CP conserving scenario, regardless of the value of the \textit{vev}-phase $\alpha$.

Furthermore, in this parameter space region, the mixing angle $\theta_{23}$ falls within the lower octant ($\theta_{23}<45^o$) for the normal hierarchy (NH) and the upper octant ($\theta_{23}>45^o$) for the inverted hierarchy (IH), as illustrated in Fig. \ref{fig3}. Specifically, we note that the mixing angle $\theta_{23}$ is approximately $44^o$ for the normal hierarchy and $47^o$ for the inverted hierarchy. In the subsequent section, we will explore an alternative scenario by increasing the \textit{vev} $v_1$ to observe its impact on CP violation in the leptonic sector.

\subsection{When \textit{vev} $v_1$ is in the GeV range}
 
In this specific scenario, we randomly varied the \textit{vev} $v_1$ within the range of 10 to 17 GeV to analyze the influence of both \textit{vevs} on the parameter space. Once again, in this case, $v_2$ is determined by the relation $v_2 = \sqrt{v^2-v_1^2}$ GeV. The predictions obtained for masses and other parameters will be discussed as follows.

\subsubsection{Charged lepton masses}

In this case we have randomly varied the Yukawa couplings in range as shown in third column of Table \ref{data} and after diagonalizing the mass matrix $M_lM_l^\dagger$ numerically using Eqn.(\ref{mld}) we have obtained squared-masses of the charged leptons. The benchmark point and prediction for the masses of charged leptons are listed in third column of Tables \ref{data3} and \ref{data5}, respectively.

\subsubsection{Neutrino Oscillation Parameters}

The ranges of Yukawa couplings considered in the numerical analysis are presented in the third column of Table \ref{data2}. In both normal and inverted hierarchies of neutrinos, we have identified the benchmark point and obtained the values of the mixing angles and mass squared differences, as shown in the third column of Tables \ref{data4} and \ref{data6}, respectively.

For these values, we have determined that $|m_{ee}|$ is approximately 0.01191 eV for NH and 0.04726 eV for IH. Even with the increased value of $v_1$, the masses $m_1$, $m_2$, and $m_3$ follow the same trend as in the previous case when $v_1 << v_2$.

The correlation plots, in Fig. \ref{fig5}, are shown for NH (first row) and IH (second row) cases. It can be seen from Fig. \ref{fig5} that $|m_{ee}|$ is strongly correlated to $m_1$ (Fig. \ref{fig5}(a)) and Jarlskog invariant $J_{CP}\in(-0.006\rightarrow0.006)$ (Fig. \ref{fig5}(b)) for NH. For IH $|m_{ee}|\in (4.67\rightarrow4.83)\times 10^{-2}$ eV (Fig. \ref{fig5}(c)) and $J_{CP}\in (-0.021\rightarrow-0.018)\oplus (0.018\rightarrow0.021)$ (Fig. \ref{fig5}(d)). It is interesting to note that in case of IH, the CP violating phase $\delta=0$ is disallowed thus making this scenario necessarily CP violating. Another characteristic feature of the model is predicted correlation between yet unknown $\theta_{23}$-octant and neutrino mass hierarchy. In Fig. \ref{fig6} we have shown allowed parameter space of the model in ($\theta_{13}-\theta_{23}$) plane. It is evident from these plots that, for NH (IH),  mixing angles ($\theta_{13}-\theta_{23}$) exhibit negative (positive) correlation. Thus, $\theta_{23}$ resides in the lower octant for NH and in higher octant for IH see Fig. \ref{fig6}. Further, the model exhibit a sharp prediction for $\theta_{23}$ approximately equal to $43^o$ ($48^o$) for NH (IH). Similarly, $\theta_{13}$ is found to be around $8.8^o$ ($8.24^o$) for NH (IH). Some other interesting correlation plots are shown in Fig. \ref{fig7}.

Multi-Higgs models often lead to significant flavor-changing neutral currents (FCNC)\cite{Crivellin:2013wna}. One approach to reduce these couplings is by aligning all right-handed fermions to interact with a single Higgs. This alignment can be achieved through an additional global $Z_2$ symmetry\cite{Glashow:1976nt,Paschos:1976ay}. Alternatively, it can result from adjusting the ratio of the \textit{vevs} $v_2/v_1$. In our study, we are investigating the consequences of Generalized
CP (GCP) symmetry without imposing additional symmetry constraints on the most
general Yukawa couplings and their corresponding neutrino phenomenology. Within the model it is possible that the Higgs responsible for FCNC is very heavy and thus, produce vanishing FCNCs. The examination of FCNC effects in the current model is beyond the scope of this study and will be explored in future research.

\section{Conclusions}

In extended theoretical frameworks beyond the Standard Model, the count of free parameters increases compared to those at low energy. By introducing additional symmetry into the Lagrangian, we can substantially reduce the number of free parameters. In this study, we investigate the impact of GCP symmetries in the leptonic sector within the context of the 2HDM model. To incorporate non-zero neutrino masses, we extend the lepton sector by introducing right-handed neutrinos through the Type-I seesaw mechanism.

The scalar potential of the 2HDM model typically involves fourteen free parameters. However, due to the GCP symmetry we impose, this number reduces to four in the unbroken CP3 case and six in the softly broken CP3 symmetry case. Consequent to GCP, the charged lepton (Eqn. (\ref{gs})) and neutrino Yukawa coupling (Eqn. (\ref{ys})) matrices contain 12 independent parameters, six each in charged lepton and  neutrino sectors. Also, Majorana mass term have two real parameters (Eqn. (\ref{mr})).

This model exhibits a rich phenomenology and reveals strong correlations among neutrino oscillation parameters. The complex \textit{vev} phase $\alpha$ is the sole source of CP violation in the model. We consider two distinct phenomenological scenarios:
\begin{enumerate}
    \item In the first scenario, where \textit{vev} $v_1$ is much smaller than $v_2$, CP is conserved regardless of the value of the \textit{vev}-phase $\alpha$ (see Fig. \ref{fig2}). This scenario provides a unique phenomenology for normal and inverted hierarchies of neutrinos. Notably, the model precisely predicts the neutrino mixing angles, particularly the atmospheric mixing angle $\theta_{23}$.
    \item In the second scenario, where $v_1$ is in the GeV range, the atmospheric mixing angle $\theta_{23}$ is below (for NH) or above (for IH) maximality, approximately $\approx43^o$ and $\approx48^o$ respectively (see Fig. \ref{fig6}). The Dirac CP phase $\delta$ is tightly constrained to be within the range of $-10^o$ to $10^o$ for the NH case. However, if the neutrino masses follow an inverted mass spectrum, the model inherently exhibits CP violation with $\delta$ approximately equal to $\pm 40^o$.

In summary, our investigation into the 2HDM model with GCP symmetries reduces the number of free parameters, leading to precise predictions for neutrino mixing angles and distinct CP-violating scenarios, shedding light on its unique phenomenology.

\end{enumerate}

\noindent\textbf{\Large{Acknowledgments}}
 \vspace{.3cm}\\
Tapender acknowledges the financial support provided by Central University of Himachal Pradesh. The authors, also, acknowledge Department of Physics and Astronomical Science for providing necessary facility to carry out this work.



\begin{thebibliography}{100}
\bibitem{Minkowski:1977sc}
P.~Minkowski,
Phys. Lett. B \textbf{67}, 421-428 (1977).

\bibitem{Yanagida:1979as}
T.~Yanagida,
Conf. Proc. C \textbf{7902131}, 95-99 (1979).

\bibitem{Glashow:1979nm}
S.~L.~Glashow,
NATO Sci. Ser. B \textbf{61}, 687 (1980).

\bibitem{Gell-Mann:1979vob}
M.~Gell-Mann, P.~Ramond and R.~Slansky,
Conf. Proc. C \textbf{790927}, 315-321 (1979).

\bibitem{Mohapatra:1979ia}
R.~N.~Mohapatra and G.~Senjanovic,
Phys. Rev. Lett. \textbf{44}, 912 (1980).

\bibitem{Lee:1973iz}
T.~D.~Lee,
Phys. Rev. D \textbf{8}, 1226-1239 (1973).

\bibitem{Castano:1993ri}
D.~J.~Castano, E.~J.~Piard and P.~Ramond,
Phys. Rev. D \textbf{49}, 4882-4901 (1994).

\bibitem{Weinberg:1977ma}
S.~Weinberg,
Phys. Rev. Lett. \textbf{40}, 223-226 (1978).


\bibitem{Wilczek:1977pj}
F.~Wilczek,
Phys. Rev. Lett. \textbf{40}, 279-282 (1978).

\bibitem{ParticleDataGroup:2022pth}
R.~L.~Workman \textit{et al.} [Particle Data Group],
PTEP \textbf{2022}, 083C01 (2022).

\bibitem{Gunion:2002zf}
J.~F.~Gunion and H.~E.~Haber,
Phys. Rev. D \textbf{67}, 075019 (2003).

\bibitem{Glashow:1976nt}
S.~L.~Glashow and S.~Weinberg,
Phys. Rev. D \textbf{15}, 1958 (1977).

\bibitem{Paschos:1976ay}
E.~A.~Paschos,
Phys. Rev. D \textbf{15}, 1966 (1977).

\bibitem{Peccei:1977hh}
R.~D.~Peccei and H.~R.~Quinn,
Phys. Rev. Lett. \textbf{38}, 1440-1443 (1977).

























\bibitem{Antusch:2001vn}
S.~Antusch, M.~Drees, J.~Kersten, M.~Lindner and M.~Ratz,
Phys. Lett. B \textbf{525}, 130-134 (2002).

\bibitem{Atwood:2005bf}
D.~Atwood, S.~Bar-Shalom and A.~Soni,
Phys. Lett. B \textbf{635}, 112-117 (2006).

\bibitem{Clarke:2015hta}
J.~D.~Clarke, R.~Foot and R.~R.~Volkas,
Phys. Rev. D \textbf{92}, no.3, 033006 (2015).

\bibitem{Liu:2016mpf}
Z.~Liu and P.~H.~Gu,
Nucl. Phys. B \textbf{915}, 206-223 (2017).

\bibitem{Camargo:2019ukv}
D.~A.~Camargo, M.~D.~Campos, T.~B.~de Melo and F.~S.~Queiroz,
Phys. Lett. B \textbf{795}, 319-326 (2019).

\bibitem{Cogollo:2019mbd}
D.~Cogollo, R.~D.~Matheus, T.~B.~de Melo and F.~S.~Queiroz,
Phys. Lett. B \textbf{797}, 134813 (2019).

\bibitem{LopezHonorez:2006gr}
L.~Lopez Honorez, E.~Nezri, J.~F.~Oliver and M.~H.~G.~Tytgat,
JCAP \textbf{02}, 028 (2007).

\bibitem{Gustafsson:2007pc}
M.~Gustafsson, E.~Lundstrom, L.~Bergstrom and J.~Edsjo,
Phys. Rev. Lett. \textbf{99}, 041301 (2007).

\bibitem{Dolle:2009fn}
E.~M.~Dolle and S.~Su,
Phys. Rev. D \textbf{80}, 055012 (2009).

\bibitem{LopezHonorez:2010eeh}
L.~Lopez Honorez and C.~E.~Yaguna,
JHEP \textbf{09}, 046 (2010).

\bibitem{Arcadi:2018pfo}
G.~Arcadi,
Eur. Phys. J. C \textbf{78}, no.10, 864 (2018).


\bibitem{Bertuzzo:2018ftf}
E.~Bertuzzo, S.~Jana, P.~A.~N.~Machado and R.~Zukanovich Funchal,
Phys. Lett. B \textbf{791}, 210-214 (2019).




\bibitem{Gunion:2005ja}
J.~F.~Gunion and H.~E.~Haber,
Phys. Rev. D \textbf{72}, 095002 (2005).

\bibitem{Ferreira:2009wh}
P.~M.~Ferreira, H.~E.~Haber and J.~P.~Silva,
Phys. Rev. D \textbf{79}, 116004 (2009).

\bibitem{Ferreira:2010bm}
P.~M.~Ferreira and J.~P.~Silva,
Eur. Phys. J. C \textbf{69}, 45-52 (2010).





\bibitem{Velhinho:1994np}
J.~Velhinho, R.~Santos and A.~Barroso,
Phys. Lett. B \textbf{322}, 213-218 (1994).


\bibitem{Ivanov:2006yq}
I.~P.~Ivanov,
Phys. Rev. D \textbf{75}, 035001 (2007)
[erratum: Phys. Rev. D \textbf{76}, 039902 (2007)].

\bibitem{Maniatis:2006fs}
M.~Maniatis, A.~von Manteuffel, O.~Nachtmann and F.~Nagel,
Eur. Phys. J. C \textbf{48}, 805-823 (2006).

\bibitem{Nishi:2006tg}
C.~C.~Nishi,
Phys. Rev. D \textbf{74}, 036003 (2006)
[erratum: Phys. Rev. D \textbf{76}, 119901 (2007)].

\bibitem{Ivanov:2007de}
I.~P.~Ivanov,
Phys. Rev. D \textbf{77}, 015017 (2008).


\bibitem{Maniatis:2007vn}
M.~Maniatis, A.~von Manteuffel and O.~Nachtmann,
Eur. Phys. J. C \textbf{57}, 719-738 (2008).



















\bibitem{Ecker:1987qp}
G.~Ecker, W.~Grimus and H.~Neufeld,
J. Phys. A \textbf{20}, L807 (1987).





\bibitem{deSalas:2020pgw}
P.~F.~de Salas, D.~V.~Forero, S.~Gariazzo, P.~Mart\'\i{}nez-Mirav\'e, O.~Mena, C.~A.~Ternes, M.~T\'ortola and J.~W.~F.~Valle,
JHEP \textbf{02}, 071 (2021).

\bibitem{Jarlskog:1985ht}
C.~Jarlskog,
Phys. Rev. Lett. \textbf{55}, 1039 (1985).

\bibitem{Bilenky:1987ty}
S.~M.~Bilenky and S.~T.~Petcov,
Rev. Mod. Phys. \textbf{59}, 671 (1987)
[erratum: Rev. Mod. Phys. \textbf{61}, 169 (1989); erratum: Rev. Mod. Phys. \textbf{60}, 575-575 (1988)].

\bibitem{Krastev:1988yu}
P.~I.~Krastev and S.~T.~Petcov,
Phys. Lett. B \textbf{205}, 84-92 (1988).

\bibitem{Crivellin:2013wna}
A.~Crivellin, A.~Kokulu and C.~Greub,
Phys. Rev. D \textbf{87}, no.9, 094031 (2013).



\end{thebibliography}
\end{document}